# Nonreductive WZW models and their CFTs

José M. Figueroa-O'Farrill[†]

*Department of Physics, Queen Mary and Westfield College*
*Mile End Road, London E1 4NS, UK*

and

Sonia Stanciu[‡]

*ICTP, P.O. Box 586, I-34100 Trieste, ITALY*

### Abstract

We study two-dimensional WZW models with target space a nonreductive Lie group. Such models exist whenever the Lie group possesses a bi-invariant metric. We show that such WZW models provide a lagrangian description of the nonreductive (affine) Sugawara construction. We investigate the gauged WZW models and we prove that gauging a diagonal subgroup results in a conformal field theory which can be identified with a coset construction. A large class of exact four-dimensional string backgrounds arise in this fashion. We then study the topological conformal field theory resulting from the $G/G$ coset. We identify the Kazama algebra extending the BRST algebra, and the BV algebra structure in BRST cohomology which it induces.

[†] e-mail: j.m.figueroa@qmw.ac.uk

[‡] e-mail: sonia@ictp.trieste.it

§1   INTRODUCTION

WZW models (both "as is" or gauged) have long been used as exactly solvable models for string backgrounds, and some even describe realistic spacetime solutions to string theory. For a $\sigma$-model to define a consistent string background it has to be quantum-mechanically conformally invariant: in other words, its $\beta$ function must vanish. In the particular case of the WZW model, quantum conformal invariance is guaranteed by the affine[1] Sugawara construction. This fact makes them prime ingredients in building exact string backgrounds. Since the standard Sugawara construction uses the Killing form on the Lie algebra, one may fear that in the nonsemisimple case (in which the Killing form is necessarily degenerate) the Sugawara construction does not exist and hence that WZW models with target space a nonsemisimple Lie group do not yield possible string backgrounds. However after the work of Nappi and Witten [1] we now know that this is not necessarily the case. Indeed, in [1] a exact string background was constructed using a WZW model with target a solvable four-dimensional Lie group: the centrally-extended Poincaré group in two-dimensions. This background is an exact solution to the $\beta$-function equation and describes a gravitational plane wave in four-dimensions.

This result sparked much interest in WZW models based on nonsemisimple Lie groups, and along with this, in Sugawara constructions based on nonsemisimple Lie algebras [2] [3] [4] [5] [6] [7] [8] [9] [10]. The condition for the existence of a Sugawara construction starting from a general WZW model was given by Mohammedi in [11] (see also [12] and [13]). The conclusion of [11] and [13] is that the Sugawara construction exists whenever the target Lie group of the WZW model admits a bi-invariant metric. Lie groups possessing a bi-invariant metric include the reductive Lie groups (which include the compact Lie groups) but also more exotic Lie groups like the one in the WZW model of [1]. Although Lie groups with an invariant metric are not classified (but see [14] for a start), there exists a structure theorem [15] for their Lie algebras which tells us how to construct them. In [13] we exploited this theorem to prove that the general Sugawara construction always factorises into a semisimple construction and a nonsemisimple construction, in such a way that the central charge of the nonsemisimple construction is always equal to the dimension of the corresponding Lie algebra, and thus an integer. In particular, this limits the ways in which four-dimensional string backgrounds can be constructed from nonreductive WZW models. In fact, as shown later in [14], there are only two WZW models which give rise to exact four-dimensional string backgrounds: the one in [1] and the one in [5]. Therefore to construct new string backgrounds out of WZW data one is forced to consider gauged WZW models and, in particular, their cosets.

---

[1]   In this paper we shall refer to the affine Sugawara construction simply as the Sugawara construction, hoping that no confusion shall arise in the process.

The motivation of the present paper, which in some ways can be considered as an extension of [**13**], is therefore twofold: first, to understand the precise relation between the WZW model and the Sugawara construction in the nonreductive case; and having achieved this, to define new coset conformal field theories by gauging nonreductive WZW models. Hopefully this then opens the possibility of searching for new (four-dimensional) exact string backgrounds. Several examples of gauged nonreductive WZW models have already been considered in the literature [**16**] [**14**] [**17**], which can be considered special cases of the ones we discuss in this paper. In this paper we attempt to convince the reader that the study of self-dual Lie algebras (Lie algebras with an invariant metric) is important towards the construction of exact string backgrounds starting from WZW models. Lack of spacetime forbids that we devote time to the detailed investigation of the properties of any given example; but we do point out the existence of a large class of new four-dimensional exact string backgrounds based on the solvable Lie algebras considered in [**18**].

We now describe the contents of this paper. We shall show that nonreductive WZW models still provide a lagrangian description of the Sugawara construction and that certain gauged WZW models provide a lagrangian description of the coset construction—thus extending well-known results to the nonreductive case. This proves the quantum conformal invariance of a large class of $\sigma$-models with target space nonreductive Lie groups and their homogeneous spaces; and makes them prime candidates from which to construct new exact string backgrounds. We shall point to a large class of new examples when we discuss the general coset construction. We hasten to add that contrary to the reductive case, these results are not immediate. They follow, as we will see, after some detailed knowledge of the structure of nonreductive self-dual Lie algebras.

As we will review in Section 2, the necessary and sufficient condition for the existence of a WZW model with target space a Lie group $G$, is that $G$ should admit a bi-invariant metric $g$. This is the metric which enters in the classical action and in the classical expression for the energy-momentum tensor. Moreover it is this tensor which appears in both the classical and quantum algebra of currents. Similarly, as we will also review in Section 2, for the current algebra of a WZW model to admit a Sugawara construction one requires that $\frac{1}{2}\Omega = g + \frac{1}{2}\kappa$ should be nondegenerate [**11**] [**13**], where $\kappa$ is the Killing form. Hence the WZW model will provide a lagrangian description of the Sugawara construction if and only if both $g$ and $\Omega$ are nondegenerate. This should be familiar from the case when $G$ is simple—although since on such a $G$ there is only one conformal class of bi-invariant metrics, both $g$ and $\Omega$ are proportional to $\kappa$ and one only perceives a "renormalisation" of the expression for the quantum-mechanical energy-momentum tensor, and nondegeneracy is not an issue unless one is in either of the two pathological points: $g = 0$—which corresponds to level zero WZW model, where the classical action vanishes, or $g = -\frac{1}{2}\kappa$—which corresponds to the critical level, where the Sugawara construction breaks down. The case $G$ abelian corresponds to free bosons, and there we see that $g$ and $\Omega$ agree, and they are both nondegenerate. On the other hand, when $G$ is nonreductive, it is not at all clear that both conditions can be satisfied.

This point was glossed over in [**11**] and until now this had been a puzzle which seemed to cast a shadow of doubt on the exact nature of the relationship between the WZW model and the Sugawara construction in the nonreductive case. But as we will see in Section 3, if $g$ is nondegenerate, so will be $\Omega$ and viceversa (excluding the pathological cases alluded to above). As a consequence, the WZW model still provides a lagrangian description of the Sugawara construction, even in the nonreductive case. To prove this result it turns out to be necessary to use some knowledge about the structure of those Lie groups admitting a bi-invariant metric. In the main body of the paper we review very briefly what we need; but the interested reader is encouraged to look up the details in [**19**] where they are worked out.

We then fix once and for all a Lie group $G$ and a bi-invariant metric on it, and we then turn our attention to gauging the corresponding WZW model. As in the reductive case, there will be obstructions to gauging a subgroup $H \subset G \times G$ of the isometry group; but they are overcome, in particular, if we choose $H$ diagonal. Thus in Section 4 we consider in detail the diagonal (vector) gauging of such a WZW model. After some manipulations in the functional integral, we arrive to a conformal field theory which we treat in Section 5 in the BRST formalism. We show that the resulting theory is conformal, with energy-momentum tensor given by a nonreductive coset construction, which we previously define. This allows us to conclude that the lagrangian interpretation of the coset construction in terms of a gauged WZW model [**20**] does survive in the nonreductive case. We also give some examples of new four-dimensional string backgrounds based on the WZW models constructed in [**18**].

In Section 6 we turn our attention to the $G/G$ topological conformal field theory. We show that the BRST complex defining the theory has a natural Kazama algebra structure which embeds the BRST algebra and which induces certain operations on the BRST cohomology. In turn, these operations makes the resulting TCFT into a Batalin-Vilkowisky algebra. This is proven for generic TCFTs induced from Kazama algebras, not necessarily those coming from the $G/G$ gauged WZW model.

Finally, in Section 7 we conclude the paper with a brief summary of the main results, some comments on the obvious supersymmetric extensions of these results, and some further results we hope to obtain in future work or have already obtained and will appear elsewhere.

§2 THE NONREDUCTIVE WZW MODEL

In this section we introduce the WZW model associated with a Lie group with a bi-invariant metric. We derive the classical and quantum algebras of currents, and we review the corresponding Sugawara construction.

The WZW model

The two-dimensional WZW model is a classical conformally invariant field theory whose basic fields are harmonic maps from a Riemann surface to a Lie group. Let us therefore consider $(\Sigma, \rho)$ to be an orientable Riemann surface and for the group $G$ we will take a connected, simply connected, *not necessarily semisimple* Lie group, whose Lie algebra we will denote by $\mathfrak{g}$. Notice that since we are interested in nonreductive groups, we will *not* take $G$ to be compact, for all compact Lie groups are reductive.

Given a basis $\{X_a\}$ for $\mathfrak{g}$, we denote by $\tilde{X}_a$ the corresponding left-invariant vector fields. Then any invariant metric $\langle -,- \rangle$ on $\mathfrak{g}$ induces a bi-invariant metric on $G$, which we denote in the same way, and which is defined by

$$\left\langle \tilde{X}_a, \tilde{X}_b \right\rangle = \langle X_a, X_b \rangle . \tag{2.1}$$

Further, we denote by

$$\theta = \tilde{\theta}^a X_a , \tag{2.2}$$

the left-invariant Maurer-Cartan form on $G$, with $\tilde{\theta}^a$ being the corresponding left-invariant one-forms dual to $\tilde{X}_a$.

If we now consider maps $g : \Sigma \to G$, we can pull the Maurer-Cartan form back on $\Sigma$ to obtain $g^*\theta \in \Omega^1(\Sigma) \otimes \mathfrak{g}$, and analogously we can also pull back the $G$ metric on $\Sigma$. If we take, without loss of generality, $G$ to be a matrix group we can write $g^*\theta = g^{-1}dg$. In particular for our two-dimensional $\Sigma$

$$g^*\theta = g^{-1}\partial g + g^{-1}\bar{\partial}g . \tag{2.3}$$

Classically the WZW model is defined by an action consisting of two terms with suitably chosen coefficients,

$$I[g] = \alpha I_{kin}[g] + \beta I_{WZ}[\tilde{g}] ,\qquad(2.4)$$

which we will describe now separately. The first term

$$I_{kin}[g] = \tfrac{1}{2}||d\,(g^*\theta)\,||^2 = \tfrac{1}{2}\int_\Sigma \langle g^*\theta\,,\star(g^*\theta)\rangle ,\qquad(2.5)$$

is a kinetic term, which appears to depend on the metric on $\Sigma$ because of the Hodge star; however in two-dimensions and acting on 1-forms, $\star^2 = -1$, whence it is a complex structure and depends only on the conformal class of the metric. This renders $I_{kin}[g]$ conformally invariant. Due to this fact, and using (2.3), one can rewrite $I_{kin}$ in the more familiar form

$$I_{kin}[g] = i \int_\Sigma \langle g^{-1}\partial g\,,\,g^{-1}\bar\partial g\rangle .\qquad(2.6)$$

We will assume that the metric $\langle -\,,\,-\rangle$ is nondegenerate, for if this were not the case, the theory would be constrained and would not describe string propagation on $G$, but presumably on some quotient group.

The second term—the Wess-Zumino (WZ) term—is a nonlocal term defined on a three-dimensional manifold $B$, with boundary $\partial B = \Sigma$. This means that $g$ has to be extended to a map $\tilde{g} : B \to G$ such that $\tilde{g}|_{\partial B=\Sigma} = g$. The WZ term will thus be defined by

$$I_{WZ}[\tilde{g}] = \int_B \tilde{g}^* H = \int_B \langle \tilde{g}^*\theta\,,\,d(\tilde{g}^*\theta)\rangle .\qquad(2.7)$$

A few remarks are in order. First of all notice that not only is $\tilde{g}^* H$ (trivially) a closed form on $B$, but also $H$ itself is a closed form on $G$. Also notice that the WZ term does not depend on the metric on $\Sigma$. Although it appears to depend on the extension $\tilde{g}$, it will finally yield local equations of motion, as one can easily see by computing the variation of the WZ term:

$$\delta I_{WZ}[g] = 3 \int_\Sigma \langle g^{-1}\delta g\,,\,d(g^{-1}dg)\rangle .\qquad(2.8)$$

The dependence on $\tilde{g}$ will have no repercussions at the quantum level (that is, in the path integral) provided the WZ term is suitably normalised. Indeed, if $H$ is a nontrivial de Rham class in $G$, this imposes an integrality condition on the cohomology class of $H$ in $H^*(G)$.

The only thing that remains to be specified is the metric on $\mathfrak{g}$, and with this we arrive to the central theme of this paper. It is clear from the preceding discussion that all we need in order to define the WZW action is a Lie group which possesses a bi-invariant metric or, equivalently, a Lie algebra with an invariant metric

$$\langle X_a , X_b \rangle = \gamma_{ab} \ . \tag{2.9}$$

From now on, we will refer to such Lie algebras as *self-dual*. In the most familiar case of a WZW model defined on a simple Lie group, the metric in $\mathfrak{g}$ is induced from the trace in some faithful representation. This simplification is due to the fact that in this case there is (up to a multiplicative constant) a unique invariant metric on $\mathfrak{g}$—the Killing form,

$$\kappa_{ab} = \operatorname{Tr} \operatorname{ad} X_a \operatorname{ad} X_b \ . \tag{2.10}$$

Nevertheless, in the general case, the metric is not given by the Killing form since this will only be nondegenerate in the semisimple case, nor will it be possible generally to induce the metric from the trace in some representation.

The current algebra

The WZW model is defined by the classical action

$$\alpha^{-1} I_\gamma[g] = \tfrac{1}{2} \int_\Sigma \langle g^{-1}dg , g^{-1} \star dg \rangle - \tfrac{i}{3} \int_B \langle \tilde{g}^{-1}d\tilde{g} , d(\tilde{g}^{-1}d\tilde{g}) \rangle \ , \tag{2.11}$$

where the subscript $\gamma$ stands from now on for the metric we use in order to define the model. The coefficients of the two terms in the action have been chosen for reasons which are standard [21]. The quantum field theory will be as usual described by the path integral

$$Z = \int [dg] e^{-I_\gamma[g]} \ . \tag{2.12}$$

Independence of the quantum theory on the extension $\tilde{g}$ will in general quantise $\alpha$.

As we discussed above $I_\gamma[g]$ is conformally invariant, but it is also invariant under the group of isometries of the metric; since the metric is bi-invariant, the isometry group is $G \times G$. Yet the most important feature of the action (2.11)—one which would not be true had the relative coefficients not been chosen this

way—is its invariance under the infinite-dimensional group $G(z) \times G(\bar{z})$ acting by

$$g(z, \bar{z}) \mapsto \Omega^{-1}(z) g(z, \bar{z}) \bar{\Omega}(\bar{z}) , \qquad (2.13)$$

with $\Omega$ and $\bar{\Omega}$ being holomorphic and antiholomorphic maps, respectively, from $\Sigma$ to the group $G$. This invariance is characterised by the conserved currents

$$J(z) = -2i\alpha \partial g g^{-1}, \qquad \bar{J}(\bar{z}) = 2i\alpha g^{-1} \bar{\partial} g , \qquad (2.14)$$

with the equations of motion $\bar{\partial} J = \partial \bar{J} = 0$. If we take $J$ and $\bar{J}$ as the dynamical variables and compute the fundamental Poisson brackets we get

$$\{J_a(z), J_b(w)\} = (2i\alpha \gamma_{ab} \partial_w + f_{ab}{}^c J_c(w)) \delta(z - w) , \qquad (2.15)$$

which yields upon quantisation

$$J_a(z) J_b(w) = \frac{2i\alpha \gamma_{ab}}{(z-w)^2} + \frac{f_{ab}{}^c J_c(w)}{z-w} + \text{reg} , \qquad (2.16)$$

(and similar formulas for $\bar{J}$). Hence the (modes of the) currents satisfy an affine algebra $\widehat{\mathfrak{g}}$, whose central extension is defined by the metric of the WZW action. As it is well-known, the non-perturbative proof of the conformal invariance of the WZW model relies on the Sugawara construction based on the current algebra (2.16), which we now review.

The Sugawara construction

Let us first rescale the metric in (2.16) and assume that our currents obey the current algebra

$$J_a(z) J_b(w) = \frac{g_{ab}}{(z-w)^2} + \frac{f_{ab}{}^c J_c(w)}{z-w} + \text{reg} , \qquad (2.17)$$

By a Sugawara construction we mean the construction of a Virasoro algebra out of (normal ordered) bilinears in the currents $J_a(z)$, with the property that the currents are primary fields of conformal weight one. We take therefore as a general Ansatz for the energy-momentum tensor:

$$T(z) = \Omega^{ab} (J_a J_b)(z) , \qquad (2.18)$$

with $\Omega^{ab}$ a yet unspecified symmetric bivector. If we now impose that the currents $J_a(z)$ be primary fields of conformal weight one

$$J_a(z)T(w) = \frac{J_a(w)}{(z-w)^2} + \text{reg} ,\qquad(2.19)$$

we obtain that $\Omega^{ab}$ must be invariant and satisfy the following relation

$$2g_{ac}\Omega^{cb} + f_{ac}{}^d f_{de}{}^b \Omega^{ce} = \delta_a{}^b .\qquad(2.20)$$

This in turn, can be easily shown to be equivalent to $\Omega^{ab}$ being invertible, with inverse given by

$$\Omega_{ab} = 2g_{ab} + \kappa_{ab} ,\qquad(2.21)$$

where $\kappa$ is the Killing form of $\mathfrak{g}$ given by (2.10). A short calculation will then show that the energy-momentum tensor (2.18) obeys the Virasoro algebra with central charge

$$c = 2g_{ab}\Omega^{ab} = \dim \mathfrak{g} - \kappa_{ab}\Omega^{ab} .\qquad(2.22)$$

It was shown in [13] that the general Sugawara construction based on a self-dual Lie algebra $\mathfrak{g}$ factorized into a semisimple construction and a nonsemisimple construction, with the property that the central charge arising from the nonsemisimple Sugawara construction was always an integer.

Summarising, we have seen that for the Sugawara construction to exist it is necessary and sufficient that $\Omega_{ab}$ given by (2.21) be invertible; whereas it is $g_{ab}$ that needs to be invertible for the WZW model to be a lagrangian realisation of the CFT given by (2.17). *Can both of these conditions be satisfied simultaneously?* If $\mathfrak{g}$ were abelian, then this is clear since the Killing form is zero. On the other extreme, if $\mathfrak{g}$ were simple, then both $\Omega_{ab}$ and $g_{ab}$ would be proportional to $\kappa_{ab}$ and except at two pathological points, both conditions would be satisfied. These two pathological points are $g_{ab} = 0$ ("zero level"), where the action itself is zero; and $\Omega_{ab} = 0$ ("critical level"), where the Sugawara construction breaks down. Similarly if $\mathfrak{g}$ were reductive—that is, the direct product of abelian and simple factors—then both conditions are satisfied except at the pathological points of the simple factors. But how about in the more general case of a self-dual Lie algebra? A detailed analysis given in the next section will show that both conditions can be simultaneously satisfied.

## §3  Nondegeneracy of $g$ and $\Omega$

In this section we will prove that for *any* self-dual Lie algebra $\mathfrak{g}$, both bilinear forms $g$ and $\Omega$ related by (2.21), are nondegenerate, except at the pathological points of its simple factors. Actually we will prove something slightly stronger, which we will state precisely after we introduce some notation that we will need in order to prove the theorem. It turns out that the proof of the above result relies on the knowledge of the structure of self-dual Lie algebras, which we now review. A more complete treatment can be found in [**19**].

The structure of self-dual Lie algebras

Although nothing like the Cartan classification for semisimple Lie algebras exists for self-dual Lie algebras, we have the next best thing: a structure theorem which tells us how they can be constructed. This structure theorem, proven by Medina and Revoy [**15**], tells us that the class of such algebras is the smallest class containing the simple and the one-dimensional Lie algebras, which is closed under the operations of taking direct sums and double extensions (see below). It is clear that if $\mathfrak{g}_1$ and $\mathfrak{g}_2$ are two Lie algebras with invariant metric, so is their direct sum $\mathfrak{g}_1 \times \mathfrak{g}_2$, with the direct sum metric. A self-dual Lie algebra that can be written as such a direct sum is called *decomposable*, and one that cannot is called *indecomposable*. The structure theorem of [**15**] can now be stated more precisely as follows:

THEOREM 3.1. *Let $\mathfrak{g}$ be an indecomposable self-dual Lie algebra. Then $\mathfrak{g}$ is either one-dimensional, simple, or else it is the double-extension of a self-dual Lie algebra by another Lie algebra which is either simple or one-dimensional.*

The operation of double extension, which follows from the proof of the above theorem, is an intriguing generalisation of a semidirect product which we now briefly describe. We consider a self-dual Lie algebra $\mathfrak{a}$, with an invariant metric $\Omega^{\mathfrak{a}}_{ij}$ relative to a fixed basis $\{X_i\}$. We consider also a Lie algebra $\mathfrak{b}$, with basis $\{H_\alpha\}$, acting on $\mathfrak{a}$ via antisymmetric derivations, together with its dual $\mathfrak{b}^*$ whose canonical dual basis we denote by $\{H^\alpha\}$. The *double extension* $\mathfrak{g} = D(\mathfrak{a}, \mathfrak{b})$ of the algebra $\mathfrak{a}$ by the algebra $\mathfrak{b}$ will be then defined on the vector space $\mathfrak{a} \oplus \mathfrak{b} \oplus \mathfrak{b}^*$ by the following Lie brackets

$$\begin{aligned}
[X_i, X_j] &= f_{ij}{}^k X_k + f_{ij\alpha} H^\alpha \ , \\
[H_\alpha, X_i] &= f_{\alpha i}{}^j X_j \ , \\
[H_\alpha, H_\beta] &= f_{\alpha\beta}{}^\gamma H_\gamma \ , \\
[H_\alpha, H^\beta] &= -f_{\alpha\gamma}{}^\beta H^\gamma \ , \\
[H^\alpha, X_i] &= [H^\alpha, H^\beta] = 0 \ ,
\end{aligned} \qquad (3.2)$$

where the action of $\mathfrak{b}$ on $\mathfrak{a}$ is reflected in the bracket $[H_\alpha, X_i]$ and where $f_{ij\alpha} = f_{\alpha i}{}^k \Omega^{\mathfrak{a}}_{kj}$. This means in particular that $D(\mathfrak{a}, \mathfrak{b})$ does not depend only on $\mathfrak{a}$ and $\mathfrak{b}$, as the notation would suggest, but also on the action of $\mathfrak{a}$ on $\mathfrak{b}$ and on the metric of $\mathfrak{a}$.

One important feature of this construction is that $\mathfrak{g}$ has an invariant metric given by

$$\Omega^{\mathfrak{g}}_{ab} = \begin{array}{c} X_i \\ H_\alpha \\ H^\alpha \end{array} \begin{pmatrix} \Omega^{\mathfrak{a}}_{ij} & 0 & 0 \\ 0 & h_{\alpha\beta} & \delta_\alpha{}^\beta \\ 0 & \delta^\alpha{}_\beta & 0 \end{pmatrix} \overset{\begin{array}{ccc} X_j & H_\beta & H^\beta \end{array}}{}, \tag{3.3}$$

where $(h_{\alpha\beta})$ is an arbitrary (possibly degenerate) invariant bilinear form in $\mathfrak{b}$. The Killing form of the above double extension $\kappa^{\mathfrak{g}}$ will of course be degenerate, having the form

$$\kappa^{\mathfrak{g}}_{ab} = \begin{array}{c} X_i \\ H_\alpha \\ H^\alpha \end{array} \begin{pmatrix} \kappa^{\mathfrak{a}}_{ij} & \kappa^{\mathfrak{g}}_{i\beta} & 0 \\ \kappa^{\mathfrak{g}}_{\alpha j} & \kappa^{\mathfrak{g}}_{\alpha\beta} & 0 \\ 0 & 0 & 0 \end{pmatrix} \overset{\begin{array}{ccc} X_j & H_\beta & H^\beta \end{array}}{}. \tag{3.4}$$

Let us mention one more fact [**13**] which will prove instrumental in what follows. If we consider the double extension $D(\mathfrak{a}, \mathfrak{b})$ of a decomposable self-dual Lie algebra $\mathfrak{a}$ having a simple factor, say $\mathfrak{a} = \mathfrak{s} \times \mathfrak{c}$, with $\mathfrak{c}$ self-dual but otherwise arbitrary, the simple factor $\mathfrak{s}$ will as well factor out of the double extension; that is,

$$D(\mathfrak{s} \times \mathfrak{c}, \mathfrak{b}) = \mathfrak{s} \times D(\mathfrak{c}, \mathfrak{b}) \ . \tag{3.5}$$

The nondegeneracy theorem

Let $(\mathfrak{g}, \Omega)$ be a self-dual Lie algebra—that is, $\mathfrak{g}$ is a Lie algebra and $\Omega$ is an invariant nondegenerate bilinear form on it—and let $\kappa$ denote its Killing form. Let $t$ be a scalar and let $g_t$ denote the bilinear form $g_t = \Omega - t\kappa$. Notice that $g_1$ is $2g$. Fix $t$ once and for all and define $\mathfrak{g}^\perp$ to be the radical of $g_t$; that is $\mathfrak{g}^\perp = \{v \in \mathfrak{g} | g_t(v, w) = 0 \ \forall w \in \mathfrak{g}\}$. Notice that $\mathfrak{g}^\perp \subset \mathfrak{g}$ is an ideal, since the bilinear form $g_t$ is invariant. In particular, $\mathfrak{g}^\perp$ is a Lie algebra. We will prove the following result:

THEOREM 3.6. *If $(\mathfrak{g}, \Omega)$ is an indecomposable self-dual Lie algebra, then $\mathfrak{g}^\perp = 0$ unless $\mathfrak{g}$ is simple and $\Omega = t\kappa$, in which case $g_t = 0$ and $\mathfrak{g}^\perp = \mathfrak{g}$.*

Notice that if $(\mathfrak{g}, \Omega)$ is indecomposable, then by the structure theorem of [15], it is either simple, one-dimensional, or a double extension $D(\mathfrak{a}, \mathfrak{b})$ where $\mathfrak{b}$ is simple or one-dimensional. Theorem 3.6 is clear for the first two cases: if $\mathfrak{g}$ is one-dimensional, then $\kappa = 0$ and $\mathfrak{g}^\perp = 0$; and similarly if $\mathfrak{g}$ is simple, then $\mathfrak{g}^\perp$ being an ideal must be either 0 or $\mathfrak{g}$; the latter case corresponding to the case $\Omega = t\kappa$. Therefore all we have left to tackle is the last case. To this effect, let $\mathfrak{g} = D(\mathfrak{a}, \mathfrak{b})$ be a double extension. Although we may restrict $\mathfrak{b}$ to be simple or one-dimensional, we will not need to. Notice that what we have to prove is that in this case $\mathfrak{g}^\perp = 0$.

This result hinges on the following crucial fact:

$$\mathfrak{g}^\perp = D(\mathfrak{a}, \mathfrak{b})^\perp \cong \mathfrak{a}^\perp , \tag{3.7}$$

where the isomorphism is one of Lie algebras. To see this, let us first see that there is a vector space isomorphism between them and then check that it is also preserves the Lie bracket. Indeed, let $v = v^j X_j + v^\alpha H_\alpha + v_\alpha H^\alpha$ belong to $\mathfrak{g}^\perp$ and let us see what this implies. The bilinear form defining $\perp$ is $\Omega^{\mathfrak{g}}_{ab} - t\kappa^{\mathfrak{g}}_{ab}$, whose matrix is given by

$$\begin{array}{c} \\ X_i \\ H_\alpha \\ H^\alpha \end{array} \begin{pmatrix} X_j & H_\beta & H^\beta \\ \Omega^{\mathfrak{a}}_{ij} - t\kappa^{\mathfrak{a}}_{ij} & -t\kappa^{\mathfrak{g}}_{i\beta} & 0 \\ -t\kappa^{\mathfrak{g}}_{\alpha j} & h_{\alpha\beta} - t\kappa^{\mathfrak{g}}_{\alpha\beta} & \delta_\alpha{}^\beta \\ 0 & \delta^\alpha{}_\beta & 0 \end{pmatrix} . \tag{3.8}$$

Therefore, $v \in \mathfrak{g}^\perp$ implies that

$$(\Omega^{\mathfrak{g}}_{ab} - t\kappa^{\mathfrak{g}}_{ab}) \begin{pmatrix} v^j \\ v^\beta \\ v_\beta \end{pmatrix} = \begin{pmatrix} (\Omega^{\mathfrak{a}}_{ij} - t\kappa^{\mathfrak{a}}_{ij})v^j - t\kappa^{\mathfrak{g}}_{i\beta}v^\beta \\ -t\kappa^{\mathfrak{g}}_{\alpha j}v^j + (h_{\alpha\beta} - t\kappa^{\mathfrak{g}}_{\alpha\beta})v^\beta + v_\alpha \\ v^\alpha \end{pmatrix} = 0 . \tag{3.9}$$

This in turn yields the equations $v^\alpha = 0$, $v_\alpha = t\kappa^{\mathfrak{g}}_{\alpha j}v^j$, and

$$(\Omega^{\mathfrak{a}}_{ij} - t\kappa^{\mathfrak{a}}_{ij})v^j = 0 , \tag{3.10}$$

whence $v^j X_j$ belongs to $\mathfrak{a}^\perp$. Conversely, any $v^j X_j \in \mathfrak{a}^\perp$ extends to a vector $v^j X_j + tv^j \kappa^{\mathfrak{g}}_{\alpha j} H^\alpha$ which by the above computation belongs to $\mathfrak{g}^\perp$. In summary, we have a vector space isomorphism $s: \mathfrak{a}^\perp \to \mathfrak{g}^\perp$, defined by $s(v^j X_j) = v^j X'_j$, where $X'_j = X_j + t\kappa^{\mathfrak{g}}_{\alpha j} H^\alpha$. We will now show that this is also an isomorphism of Lie algebras. Computing the brackets in $\mathfrak{g}$, we obtain

$$[s(X_i), s(X_j)]_{\mathfrak{g}} = [X'_i, X'_j]_{\mathfrak{g}} = f_{ij}{}^k X_k + f_{ij\alpha} H^\alpha$$

$$= f_{ij}{}^k X'_k + (f_{ij\alpha} - t f_{ij}{}^k \kappa^{\mathfrak{g}}_{k\alpha}) H^\alpha \ . \tag{3.11}$$

Now notice that $f_{ij\alpha} = f_{ij}{}^a \Omega^{\mathfrak{g}}_{a\alpha}$, and that $f_{ij}{}^k \kappa^{\mathfrak{g}}_{k\alpha} = f_{ij}{}^a \kappa^{\mathfrak{g}}_{a\alpha}$; so that we can rewrite (3.11) as

$$[X'_i, X'_j]_{\mathfrak{g}} = f_{ij}{}^k X'_k + f_{ij}{}^a (g^{\mathfrak{g}}_t)_{a\alpha} H^\alpha \ . \tag{3.12}$$

Using that $g^{\mathfrak{g}}_t$ is an invariant bilinear form, we arrive at

$$[X'_i, X'_j]_{\mathfrak{g}} = f_{ij}{}^k X'_k - f_{i\alpha}{}^a (g^{\mathfrak{g}}_t)_{aj} H^\alpha \ . \tag{3.13}$$

Finally we notice that $f_{i\alpha}{}^a (g^{\mathfrak{g}}_t)_{aj} = f_{i\alpha}{}^k (g^{\mathfrak{g}}_t)_{kj}$ and that the restriction of $g^{\mathfrak{g}}_t$ to $\mathfrak{a}$ coincides with $g^{\mathfrak{a}}_t$, so that we end up with

$$[X'_i, X'_j]_{\mathfrak{g}} = f_{ij}{}^k X'_k - f_{i\alpha}{}^k (g^{\mathfrak{g}}_t)_{jk} H^\alpha \ , \tag{3.14}$$

which shows explicitly that if $v^i X_i$ and $w^j X_j$ are in $\mathfrak{a}^\perp$, then

$$[s(v^i X_i), s(w^j X_j)]_{\mathfrak{g}} = s([v^i X_i, w^j X_j]_{\mathfrak{a}}) \ , \tag{3.15}$$

so that $s$ is a homomorphism.

We now turn to the proof of Theorem 3.6. We proceed by induction on the dimension of the Lie algebra. Suppose that the theorem is true for all indecomposable self-dual Lie algebras of dimension $\leq N$—the case $N = 1$ being trivially satisfied—and let $\mathfrak{g} = D(\mathfrak{a}, \mathfrak{b})$ be an indecomposable double-extension of dimension $N+1$. The theorem follows if we can prove that $\mathfrak{g}^\perp = 0$. Using (3.7), we have that $\mathfrak{g}^\perp \cong \mathfrak{a}^\perp$. In general $\mathfrak{a}$ need not be indecomposable, so write it as $\mathfrak{a} = \mathfrak{a}_1 \times \cdots \times \mathfrak{a}_k$, where each $\mathfrak{a}_i$ is indecomposable. Clearly, $\mathfrak{a}^\perp \cong \mathfrak{a}_1^\perp \times \cdots \times \mathfrak{a}_k^\perp$. Since $\dim \mathfrak{a}_i < \dim \mathfrak{g}$ for each $i$, we can apply the induction hypothesis to deduce that $\mathfrak{a}_i^\perp$ will only be nonzero when $\mathfrak{a}_i$ is simple. But $\mathfrak{a}$ cannot have any simple factors, for otherwise it would violate the hypothesis that $\mathfrak{g}$ is indecomposable (recall (3.5)). Therefore $\mathfrak{a}^\perp = 0$ and we can extend the hypothesis. This concludes the proof of Theorem 3.6.

As an easy corollary of our results, we notice that if $(\mathfrak{g}, \Omega)$ is any self-dual Lie algebra, then $\mathfrak{g}^\perp$ is semisimple; whence, since it is an ideal, it is a factor. Therefore, any self-dual Lie algebra $\mathfrak{g}$ decomposes into an orthogonal direct product $\mathfrak{g} = \mathfrak{g}^\perp \times \mathfrak{g}_1$, where $\mathfrak{g}^\perp$ is semisimple, and $\mathfrak{g}_1^\perp = 0$.

## §4 GAUGED WZW MODEL

Having proven that the WZW model generically provides a lagrangian realisation of the Sugawara construction, we now turn our attention to the gauged WZW model and to the CFTs that this procedure gives rise to. We will first review the conditions to be able to gauge a WZW model and we then specialise to the diagonal gauging. For this case we will be able to manipulate the functional integral and exhibit the resulting quantum field theory as a CFT whose energy-momentum tensor agrees with that of a coset construction. The analogous results in the case of $G$ reductive were obtained in [20].

Gauging the WZW model

Let us consider then the problem of gauging a subgroup $H \subset G \times G$. In other words we want to "partially" promote (2.13) to a local invariance, under transformations of the form

$$g(z,\bar{z}) \mapsto \lambda^{-1}(z,\bar{z})g(z,\bar{z})\rho(z,\bar{z}) , \qquad (4.1)$$

where $\lambda \times \rho : \Sigma \to H$ is an arbitrary smooth map. However it is not possible to gauge all symmetry groups $H$, because of some (cohomological) obstructions which have to be overcome [22] (see also the first reference of [23]). The result of this analysis is well-known. Let $H \subset G \times G$ be the subgroup of the isometry group that we are willing to gauge, and let $\mathfrak{h} \subset \mathfrak{g} \times \mathfrak{g}$ be its Lie algebra. The Lie algebra embedding $\mathfrak{h} \subset \mathfrak{g} \times \mathfrak{g}$ is characterised by two homomorphisms $\ell, r : \mathfrak{h} \to \mathfrak{g}$, so that if $X$ is an element of $\mathfrak{h}$, its image in $\mathfrak{g} \times \mathfrak{g}$ is $(\ell(X), r(X))$. Then the condition for $H$ to be gaugeable is simply that

$$\langle \ell(X), \ell(Y) \rangle = \langle r(X), r(Y) \rangle , \qquad (4.2)$$

for all $X, Y$ in $\mathfrak{h}$. The most familiar case is the one in which $\ell = r$, known as diagonal gauging, and it is this case that, for WZW models based on semisimple groups, gives rise to the coset construction. For this reason we are going to focus, in the remainder of this paper, on the case of diagonal gauging. We will see that also for $\mathfrak{g}$ self-dual, the resulting theory is conformal and we will be able to identify it with a coset construction.

Diagonal gauging

We now consider, as advertised, the problem of gauging a diagonal (and hence anomaly-free) subgroup $H \subset G \times G$, in which case our basic fields will transform according to

$$g(z, \bar{z}) \mapsto \lambda^{-1}(z, \bar{z}) g(z, \bar{z}) \lambda(z, \bar{z}) , \qquad (4.3)$$

in obvious notation. For this we have to introduce gauge fields with components $A$ and $\bar{A}$ which locally are $\mathfrak{h}^{\mathbb{C}}$-valued one forms on $\Sigma$, and which transform under the gauge transformations according to

$$\begin{aligned} A &\mapsto \lambda^{-1}(\partial + A)\lambda , \\ \bar{A} &\mapsto \lambda^{-1}(\bar{\partial} + \bar{A})\lambda . \end{aligned} \qquad (4.4)$$

By gauging the WZW model we mean constructing an extension $I_\gamma[g, A, \bar{A}]$ of (2.11) which is invariant under (4.3) and (4.4). Using the Noether procedure we obtain

$$I_\gamma[g, A, \bar{A}] = I_\gamma[g] - \int_\Sigma \langle A , \bar{J} \rangle + \langle J , \bar{A} \rangle + 2i\alpha \langle A , g^{-1}\bar{A}g \rangle - 2i\alpha \langle A , \bar{A} \rangle . \qquad (4.5)$$

Notice that, since the gauge fields have no kinetic term, they can be thought of as Lagrange multipliers: they introduce constraints at the level of the classical theory, which consist in setting the $H$-current equal to zero.

The quantum theory is described by the path integral

$$Z = \int [dg][dA][d\bar{A}] e^{-I_\gamma[g,A,\bar{A}]} . \qquad (4.6)$$

As is familiar from the study of gauge theories, we will follow the Faddeev-Popov procedure. We choose the holomorphic gauge $\bar{A} = 0$. Assuming the absence of gauge anomalies, and computing the Faddeev-Popov determinant one finds that the gauge-fixed path integral becomes

$$Z = \int [dg][dA] (\det \bar{\partial}) e^{-I_\gamma[g,A,0]} . \qquad (4.7)$$

Strictly speaking, $\bar{\partial} : C^\infty(\Sigma) \otimes \mathfrak{h}^{\mathbb{C}} \to \Omega^{0,1}(\Sigma) \otimes \mathfrak{h}^{\mathbb{C}}$ does not have a determinant since it maps to different spaces. Nevertheless, we will understand the expression $\det \bar{\partial}$ to mean the path integral

$$\det \bar{\partial} = \int [db][dc] e^{-\int_\Sigma \langle b, \bar{\partial} c \rangle} \qquad (4.8)$$

where $(b, c)$ are Faddeev-Popov ghosts which geometrically can be interpreted as follows: $c$ is an $\mathfrak{h}^{\mathbb{C}}$-valued function on $\Sigma$, and $b$ is a $(1,0)$-form on $\Sigma$ with values in the dual $(\mathfrak{h}^{\mathbb{C}})^*$, and $\langle - , - \rangle$ above indicates the dual pairing between $\mathfrak{h}^{\mathbb{C}}$ and $(\mathfrak{h}^{\mathbb{C}})^*$.

The remaining gauge fields $A$ can be parametrised as follows $A = -\partial h h^{-1}$, where $h : \Sigma \to H$ is a smooth function. We now use the celebrated Polyakov-Wiegmann identity [24], which still holds for our more general WZW models, to express $I_\gamma[g, A, 0]$ in terms of the original WZW action as follows:

$$I_\gamma[g, A, 0] = I_\gamma[gh] - I_\gamma[h] . \tag{4.9}$$

At the quantum level, the change of variables from $A$ to $h$ incurs in a jacobian factor for the functional measure of the path integral which we have to take into account. For this it is sufficient to notice that an arbitrary infinitesimal variation of the gauge field $A$ can be written as

$$\delta A = -\partial(\delta h h^{-1}) - [A, \delta h h^{-1}] = -D(\delta h h^{-1}) . \tag{4.10}$$

From this we can deduce that the jacobian of the transformation is given by

$$[dA] = (\det D) \, [dh] , \tag{4.11}$$

where $D : C^\infty(\Sigma) \otimes \mathfrak{h}^\mathbb{C} \to \Omega^{1,0}(\Sigma) \otimes \mathfrak{h}^\mathbb{C}$ denotes the holomorphic component of the covariant derivative acting on $\mathfrak{h}^\mathbb{C}$-valued functions. The same caveat as before holds for this determinant, which we represent as

$$\det D = \int [d\bar{b}][d\bar{c}] e^{-\int_\Sigma \langle \bar{b}, D\bar{c}\rangle} \tag{4.12}$$

where now $(\bar{b}, \bar{c})$ are fermionic fields (but *not* Faddeev-Popov ghosts!) with the following geometric interpretation: $\bar{c}$ is a $\mathfrak{h}^\mathbb{C}$-valued function on $\Sigma$, whereas $\bar{b}$ is a $(0, 1)$-form with values in the dual $(\mathfrak{h}^\mathbb{C})^*$. After these manipulations the path integral (4.7) becomes

$$Z = \int [dg][dh] \, (\det D) \, (\det \bar{\partial}) e^{-I_\gamma[gh]+I_\gamma[h]} , \tag{4.13}$$

where in the above expression for $D$ it is understood that $A = -\partial h \, h^{-1}$.

We will now compute the above "determinants." We will do something a little bit more general and compute

$$\det D \det \bar{D} \equiv \int [db][dc][d\bar{b}][d\bar{c}] e^{-\int_\Sigma \langle \bar{b}, D\bar{c}\rangle - \int_\Sigma \langle b, \bar{D}c\rangle} . \tag{4.14}$$

As is well-known this path-integral is determined by the integrated anomaly $W(A)$ defined by

$$\det D \det \bar{D} = e^{-W[A,\bar{A}]} \det \partial \det \bar{\partial} . \tag{4.15}$$

An explicit computation of the chiral anomaly, using a (vector) gauge invariant regularisation procedure, yields

$$\epsilon^{\mu\nu} D_\mu \left( \frac{\delta W}{\delta A^{\nu,i}(x)} \right) = \frac{1}{2\pi} \epsilon^{\mu\nu} \kappa^\mathfrak{h}_{ij} F_{\mu\nu}{}^j(x) . \tag{4.16}$$

If we are to set $A = -\partial h h^{-1}$ and $\bar{A} = -\bar{\partial}\bar{h} \, \bar{h}^{-1}$, then the integrated anomaly turns out to be nothing but a WZW action with the metric given by the Killing

– 16 –

form of $\mathfrak{h}$:
$$e^{-W(A)} = e^{I_\kappa[\bar{h}^{-1}h]} . \qquad (4.17)$$

The free chiral determinants in (4.15) we represent as in equation (4.8).

To be able to use this back in our gauged WZW model, we need to specialise to $\bar{A} = 0$ or equivalently $\bar{h} = 1$. In this fashion, we can rewrite the path integral (4.13) as follows

$$Z = \int [dg][dh][db][dc][d\bar{b}][d\bar{c}] e^{-I_\gamma[gh] + I_{\gamma+\kappa}[h]} e^{-\int \langle b, \bar{\partial} c \rangle + \langle \bar{b}, \partial \bar{c} \rangle} \qquad (4.18)$$

We now change variables $g \mapsto gh^{-1}$ in the path integral. Absence of gauge anomalies implies that the jacobian is trivial and we arrive finally at the expression

$$Z = \int [dg][dh][db][dc][d\bar{b}][d\bar{c}] e^{-I_\gamma[g] + I_{\gamma+\kappa}[h]} e^{-\int \langle b, \bar{\partial} c \rangle + \langle \bar{b}, \partial \bar{c} \rangle} \qquad (4.19)$$

Let us pause for a moment to contemplate our result. We see that we have obtained roughly the same three sectors like in the reductive case [20], with the main difference that the two WZW sectors have actions corresponding to different metrics. Furthermore one can make a similar argument to show that although the three sectors appear to be independent, there exist constraints that couple them. Basically one can gauge the vector subgroup $H$ once again in all three lagrangians, introducing *external* gauge fields and then notice that the partition function is actually independent of the gauge fields introduced, which leads to the constraint that the current which couples to this gauge field has to vanish. This current, which we will call the total current, has contributions coming from all three CFTs.

Let us consider the holomorphic sector. The total current is given by

$$J_i^{\text{tot}}(z) = J_i(z) + \tilde{J}_i(z) + J_i^{\text{gh}}(z) , \qquad (4.20)$$

where $\{J_i(z)\}$ are a subset of the $\mathfrak{g}$ currents given by (2.14), the current corresponding to the gauged sector is given by

$$\tilde{J}(z) = -\partial h h^{-1} , \qquad (4.21)$$

(where, from now on, we will drop the $2i\alpha$ factor) whereas the current corresponding to the ghost sector is defined by

$$J_i^{\text{gh}}(z) = f_{ij}{}^k (b_k c^j)(z) , \qquad (4.22)$$

with the standard point-splitting convention for the normal ordering. These currents will satisfy three commuting current algebras with the relevant OPEs

given by

$$J_i(z)J_j(w) = \frac{g_{ij}}{(z-w)^2} + \frac{f_{ij}{}^k J_k(w)}{z-w} + \text{reg} ,  \qquad (4.23)$$

$$\tilde{J}_i(z)\tilde{J}_j(w) = \frac{-g_{ij} - \kappa^{\mathfrak{h}}_{ij}}{(z-w)^2} + \frac{f_{ij}{}^k \tilde{J}_k(w)}{z-w} + \text{reg} ,  \qquad (4.24)$$

$$J_i^{\text{gh}}(z)J_j^{\text{gh}}(w) = \frac{\kappa^{\mathfrak{h}}_{ij}}{(z-w)^2} + \frac{f_{ij}{}^k J_k^{\text{gh}}(w)}{z-w} + \text{reg} .  \qquad (4.25)$$

Adding the central extensions of the three components of the total conserved current we see that they cancel each other, which just reiterates the fact that we have gauged an anomaly-free subgroup. This guarantees that the charge which generates the BRST transformations leaving the "quantum" action invariant will square to zero. The theory resulting from the gauged WZW model is then defined as the cohomology of the BRST operator: the states will be the BRST cohomology on states, and the fields will be the BRST cohomology on fields. In the case of $G$ semisimple, it is well-known that the resulting theory is again conformal—so that among the BRST invariant fields one finds an energy-momentum tensor obeying a Virasoro algebra which generates conformal transformations on all the other BRST invariant fields—and corresponds, in fact, to the CFT based on the coset $G/H$ [20] [25]. In the next section we show that something similar happens in the general gauged WZW model.

What about the anti-holomorphic sector? One might be tempted to simply repeat the above argument, but this would not be strictly speaking correct. In the holomorphic gauge $\bar{A} = 0$, the antiholomorphic "ghosts" $(\bar{b}, \bar{c})$ are *not* Faddeev-Popov ghosts since they do not arise out of gauge-fixing the action; but arise instead out of a change of variables in the path integral. In this case we do not have the right to introduce the BRST transformations in the antiholomorphic sector nor to demand that the physical states of be given by the cohomology of the BRST-type charge that we can build—even though that sector is constrained, as the argument in [20] that we have reproduced above indicates. Instead we prefer to determine the anti-holomorphic part of the theory by working in the antiholomorphic gauge $A = 0$. Of course, the end result is the same.

## §5 Coset CFTs from gauged WZW models

In this section we will analyse the quantum field theory which results after the diagonal gauging of a WZW model based on a Lie group with an invariant metric. We will prove that the resulting theory is a CFT and that it can be identified with a coset construction. Indeed, we start by analysing the coset construction in the case of self-dual Lie algebras. For the sake of simplicity we will only consider the holomorphic sector, the treatment of the antiholomorphic sector being completely analogous.

<u>The coset construction</u>

Given $(\mathfrak{g}, \Omega^{\mathfrak{g}})$ a self-dual Lie algebra, we try to construct a coset construction for a subalgebra $\mathfrak{h} \subset \mathfrak{g}$. We fix a basis $\{J_i\}$ for $\mathfrak{h}$, which we can think of as a sub-basis of the chosen basis $\{J_a\}$ for $\mathfrak{g}$. Then the OPE of the $J_i(z)$ currents is given by (the same formula as (2.17)):

$$J_i(z) J_j(w) = \frac{g_{ij}}{(z-w)^2} + \frac{f_{ij}{}^k J_k(w)}{z-w} + \text{reg} . \qquad (5.1)$$

Clearly, a coset construction exists if and only if $\mathfrak{h}$ admits a (self-dual) Sugawara construction in terms of the above currents. Naturally $\mathfrak{h}$ has to be self-dual with some invariant metric $\Omega^{\mathfrak{h}}$, but this is not all. There is an additional condition coming from the fact that the algebra of the $\mathfrak{h}$-currents is already fixed from the $\mathfrak{g}$ current algebra; that is, $\Omega^{\mathfrak{h}}$ cannot be arbitrary. Instead, it has to satisfy the following relation:

$$\Omega^{\mathfrak{h}}_{ij} - \kappa^{\mathfrak{h}}_{ij} = \Omega^{\mathfrak{g}}_{ij} - \kappa^{\mathfrak{g}}_{ij} , \qquad (5.2)$$

in an obvious notation. In other words, if $(\mathfrak{h}, \Omega^{\mathfrak{h}})$ were to admit a Sugawara construction, then the $\mathfrak{h}$-currents would have to obey an OPE such that in the second order pole one has $\frac{1}{2}(\Omega^{\mathfrak{h}}_{ij} - \kappa^{\mathfrak{h}}_{ij})$. But this is already fixed by the $\mathfrak{g}$ current algebra to be $\frac{1}{2}(\Omega^{\mathfrak{g}}_{ij} - \kappa^{\mathfrak{g}}_{ij})$. In practice, the way to test the condition (5.2) is to use (5.2) to solve for $\Omega^{\mathfrak{h}}$, and then test for its nondegeneracy. In other words, condition (5.2) is equivalent to

$$\Omega^{\mathfrak{g}}_{ij} - \kappa^{\mathfrak{g}}_{ij} + \kappa^{\mathfrak{h}}_{ij} \quad \text{is nondegenerate on } \mathfrak{h}. \qquad (5.3)$$

Notice that from the results in Section 3, condition (5.2) is roughly equivalent to saying that the restriction of $\Omega^{\mathfrak{g}} - \kappa^{\mathfrak{g}}$ to $\mathfrak{h} \subset \mathfrak{g}$ is nondegenerate. In fact, if $(\Omega^{\mathfrak{g}} - \kappa^{\mathfrak{g}})|_{\mathfrak{h}}$ is nondegenerate, then the coset construction exists except in the pathological case that $(\Omega^{\mathfrak{g}} - \kappa^{\mathfrak{g}})|_{\mathfrak{h}} = -\kappa^{\mathfrak{h}}$, in which case the $\mathfrak{h}$-currents are at the "critical level" and the Sugawara construction does not exist. Notice that this possibility only occurs when $\mathfrak{h}$ is semisimple.

Notice that if the coset construction does exist, the energy-momentum tensor

$$T_{\mathfrak{h}} = \Omega_{\mathfrak{h}}^{ij}(J_i J_j) \tag{5.4}$$

will obey the Virasoro algebra with the central charge

$$c_{\mathfrak{h}} = 2g_{ij}\Omega_{\mathfrak{h}}^{ij} = \dim \mathfrak{h} - \kappa_{ij}^{\mathfrak{h}} \Omega_{\mathfrak{h}}^{ij} \ . \tag{5.5}$$

This implies that the coset energy-momentum tensor

$$T_{\mathfrak{g}/\mathfrak{h}} \equiv T_{\mathfrak{g}} - T_{\mathfrak{h}} \tag{5.6}$$

has central charge

$$c_{\mathfrak{g}/\mathfrak{h}} = c_{\mathfrak{g}} - c_{\mathfrak{h}} = \dim \mathfrak{g} - \dim \mathfrak{h} - \left(\kappa_{ab}^{\mathfrak{g}} \Omega_{\mathfrak{g}}^{ab} - \kappa_{ij}^{\mathfrak{h}} \Omega_{\mathfrak{h}}^{ij}\right) \ . \tag{5.7}$$

Apart from the familiar examples of $\mathfrak{h} \subset \mathfrak{g}$ both reductive Lie algebras, there are some natural coset constructions among the self-dual Lie algebras. Let us take, as an example, a nonreductive indecomposable self-dual Lie algebra. By the structure theorem (Theorem 3.1) it has to be a double extension $D(\mathfrak{a}, \mathfrak{b})$ where $\mathfrak{b}$ is either simple or one-dimensional. A brief inspection at the structure of a double-extension reveals four natural subalgebras of $D(\mathfrak{a}, \mathfrak{b})$: $\mathfrak{b}$, $\mathfrak{b}^*$, $\mathfrak{b} \ltimes \mathfrak{b}^*$, and $\mathfrak{a} \ltimes \mathfrak{b}^*$. A quick glance at the invariant metric (3.3) and the Killing form (3.4) of a double extension shows that condition (5.3) is not satisfied for the subalgebras $\mathfrak{b}^*$ and $\mathfrak{a} \ltimes \mathfrak{b}^*$. How about the other two subalgebras? It turns out that in both cases, one can satisfy condition (5.3) provided that one chooses the metric on $D(\mathfrak{a}, \mathfrak{b})$ conveniently. Recall that the metric of $D(\mathfrak{a}, \mathfrak{b})$ involves an arbitrary invariant bilinear form on $\mathfrak{b}$. Then it is easy to work out that condition (5.3) can always be satisfied for a suitable choice of invariant bilinear form on $\mathfrak{b}$. In fact, it is satisfied generically. We leave the details as an exercise.

Let us simply mention the following interesting fact. If we take $\mathfrak{a}$ to be a four-dimensional abelian Lie algebra and $\mathfrak{b}$ to be one-dimensional, the double extension $D(\mathfrak{a}, \mathfrak{b})$ is a six-dimensional solvable (not necessarily nilpotent) self-dual Lie algebra. We can then coset by the subalgebra $\mathfrak{b} \ltimes \mathfrak{b}^*$ to obtain an exact four-dimensional string background. Such $D(\mathfrak{a}, \mathfrak{b})$ are not classified, but in [18] there is a classification of those whose associated $N=1$ supersymmetric WZW model admits (2,0) supersymmetry. Some explicit nonreductive coset constructions have already appeared in the literature [16] [17]. They are all special cases of the construction outlined in this paper. Lack of spacetime prevents us from discussing four-dimensional coset string backrounds more systematically; but we hope to discuss this issue elsewhere.

The CFTs in the gauged WZW model

We will now show that the gauged WZW model described in the previous section is a conformal field theory whose energy momentum tensor agrees with the coset energy-momentum tensor. We saw that the quantum field theory of the gauged WZW model is given by three quantum field theories coupled by a constraint which we can analyse in the BRST formalism. As we now show each of the three sectors of the theory is conformal, and the BRST operator preserves the total energy-momentum tensor which will be shown to be BRST-equivalent to the coset energy momentum tensor in (5.6).

We start then with the WZW CFT with group $G$ and metric $g_{ab}$. This component corresponds to the original (ungauged) WZW model which we discussed in the first part of Section 2 (recall that $g_{ab}$ is a rescaled version of $\gamma_{ab}$). There we have seen that we have a set of currents $\{J_a(z)\}_{a=1}^{\dim \mathfrak{g}}$ whose OPE is given by (2.16). We have also seen that according to the self-dual Sugawara construction this WZW sector does give rise to a CFT if the bilinear form defined in (2.21) is nondegenerate. If this is the case, the energy-momentum tensor

$$T_{\mathfrak{g}}(z) = \Omega_{\mathfrak{g}}^{ab}(J_a J_b)(z) \tag{5.8}$$

obeys a Virasoro algebra with the central charge

$$c_{\mathfrak{g}} = 2g_{ab}\Omega_{\mathfrak{g}}^{ab} = \dim \mathfrak{g} - \kappa_{ab}^{\mathfrak{g}}\Omega_{\mathfrak{g}}^{ab} \ . \tag{5.9}$$

The next ingredient is provided by the WZW model with group $H \subset G$ and metric $-(g_{ij}+\kappa_{ij}^{\mathfrak{h}})$. This is characterised by the set of currents $\{\tilde{J}_i(z)\}_{i=1}^{\dim \mathfrak{h}}$ whose OPE is given by (4.24), where $g_{ij}$ is the restriction of $g_{ab}$ to $\mathfrak{h}$. Applying (once again) the argument of Section 2, we get that this current algebra gives rise to a CFT if and only if the following bilinear form on $\mathfrak{h}$ is nondegenerate:

$$\begin{aligned}\Theta_{ij}^{\mathfrak{h}} &= -2\left(g_{ij} + \kappa_{ij}^{\mathfrak{h}}\right) + \kappa_{ij}^{\mathfrak{h}} \\ &= -\left(2g_{ij} + \kappa_{ij}^{\mathfrak{h}}\right) \ ,\end{aligned} \tag{5.10}$$

which is identical, up to the minus sign, with the metric on $\mathfrak{h}$ in the coset construction. In other words the WZW CFT corresponding to $H \subset G$ and metric $-(g_{ij}+\kappa_{ij}^{\mathfrak{h}})$ exists if and only if the self-dual Sugawara based on $\widehat{\mathfrak{h}}$ with central extension $g_{ij}$ exists. In this case the corresponding energy-momentum

tensor

$$\widetilde{T}_{\mathfrak{h}}(z) = \Theta_{\mathfrak{h}}^{ij}(\tilde{J}_i \tilde{J}_j)(z) = -\Omega_{\mathfrak{h}}^{ij}(\tilde{J}_i \tilde{J}_j)(z) , \qquad (5.11)$$

will satisfy the Virasoro algebra with central charge

$$\tilde{c}_{\mathfrak{h}} = 2\left(g_{ij} + \kappa_{ij}^{\mathfrak{h}}\right)\Omega_{\mathfrak{h}}^{ij} = \dim \mathfrak{h} + \kappa_{ij}^{\mathfrak{h}}\Omega_{\mathfrak{h}}^{ij} . \qquad (5.12)$$

The last sector of the theory consists of a set of $(\dim \mathfrak{h})$ fermionic $(b, c)$ systems of conformal weights $(1, 0)$ respectively, with OPE given by

$$b_i(z) c^j(w) = \frac{\delta_i{}^j}{z - w} + \text{reg} . \qquad (5.13)$$

The energy-momentum tensor for this $(b, c)$ system has the standard form

$$T_{\text{gh}}(z) = -\left(b_i \partial c^j\right)(z) , \qquad (5.14)$$

and obeys the Virasoro algebra with central charge

$$c_{\text{gh}} = -2\dim \mathfrak{h} . \qquad (5.15)$$

Finally, we introduce the last ingredient of this theory: the BRST operator, which we define as the contour integral of the BRST current

$$Q = \int \frac{dz}{2\pi i} j_{BRST}(z) , \qquad (5.16)$$

where the current is given by

$$j_{BRST}(z) = (c^i(J_i + \tilde{J}_i + \tfrac{1}{2}J_i^{\text{gh}}))(z) . \qquad (5.17)$$

It is an easy computation to show that $Q^2 = 0$ – which reiterates the fact that we have gauged an anomaly-free subgroup and justifies our assuming the absence of gauge anomalies.

### The energy-momentum tensor

The total energy-momentum tensor is given by the sum of the three commuting terms given by (5.8), (5.11), and (5.14):

$$T(z) = T_{\mathfrak{g}}(z) + \widetilde{T}_{\mathfrak{h}}(z) + T_{\text{gh}}(z) , \tag{5.18}$$

whose central charge is obtained by adding up (5.9), (5.12), and (5.15):

$$c = \left(\dim \mathfrak{g} - \kappa^{\mathfrak{g}}_{ab}\Omega^{ab}_{\mathfrak{g}}\right) - \left(\dim \mathfrak{h} - \kappa^{\mathfrak{h}}_{ij}\Omega^{ij}_{\mathfrak{h}}\right) . \tag{5.19}$$

Notice that this agrees with the coset central charge given by equation (5.7). This prompts us to compare $T(z)$ with the energy-momentum tensor of the corresponding coset construction. We introduce for this purpose

$$T_{\mathfrak{h}} = \Omega^{ij}_{\mathfrak{h}}(J_i J_j)(z) , \tag{5.20}$$

with $\Omega^{ij}_{\mathfrak{h}}$ given by (5.2). Our total energy-momentum tensor will then split into a sum of two *commuting* terms

$$T(z) = T_{\mathfrak{g}/\mathfrak{h}}(z) + T'(z) , \tag{5.21}$$

with $T_{\mathfrak{g}/\mathfrak{h}}(z)$ being the coset energy-momentum tensor defined by (5.6) and

$$T'(z) = T_{\mathfrak{h}}(z) + \widetilde{T}_{\mathfrak{h}}(z) + T_{\text{gh}}(z) . \tag{5.22}$$

Moreover, a short computation shows us that $T'$ satisfies a Virasoro algebra with vanishing central charge, $c' = 0$.

Our aim now is to show that the BRST charge commutes with $T$, $T_{\mathfrak{g}/\mathfrak{h}}$ and $T'$; hence they are physical operators (that is they induce operators in the physical space). Moreover we will show that $T'$ is a trivial BRST operator, which means that it induces the zero operator on physical states. For this one needs to use the following identities:

$$[Q, b_i(z)] = J^{\text{tot}}_i(z) , \tag{5.23}$$
$$[Q, c^i(z)] = -\tfrac{1}{2}f^i{}_{jk}c^j c^k(z) . \tag{5.24}$$

Using these relations we deduce the following

$$[Q, J_a(z)] = g_{aj}\partial c^j(z) - f_{aj}{}^b c^j J_b(z) , \tag{5.25}$$
$$\left[Q, \tilde{J}_i(z)\right] = -\left(g_{ij} + \kappa^{\mathfrak{h}}_{ij}\right)\partial c^j(z) - f_{ij}{}^k c^j \tilde{J}_k(z) , \tag{5.26}$$

– 23 –

$$\left[Q, J_i^{\text{gh}}(z)\right] = \kappa_{ij}^{\mathfrak{h}} \partial c^j(z) + f_{ij}{}^k c^j \left(J_k + \tilde{J}_k\right)(z) \ . \tag{5.27}$$

Finally, after a tedious but completely straightforward computation we obtain that:

$$[Q, T(z)] = \left[Q, T_{\mathfrak{g}/\mathfrak{h}}(z)\right] = \left[Q, T'(z)\right] = 0 \ ; \tag{5.28}$$

that is, $T(z)$, $T_{\mathfrak{g}/\mathfrak{h}}(z)$ and $T'(z)$ are BRST invariant. Furthermore, it also follows that there exists an operator

$$O(z) = \Omega_{\mathfrak{h}}^{ij} b_i \left(J_j - \tilde{J}_j\right)(z) \ , \tag{5.29}$$

such that

$$T'(z) = [Q, O(z)] \ ; \tag{5.30}$$

in other words, $T'(z)$ is BRST trivial, whence it is zero in cohomology.

These results imply that the quantum field theory defined by the gauged WZW model (4.19) is conformal, with energy-momentum tensor given by $T(z)$ and its antiholomorphic counterpart $\bar{T}(\bar{z})$. Moreover $T(z)$ (and similarly for $\bar{T}(\bar{z})$) is precisely the coset energy-momentum tensor $T_{\mathfrak{g}/\mathfrak{h}}(z)$; whence we conclude that the gauged WZW model provides a lagrangian realisation to the coset CFT $G/H$.

## §6  The $G/G$ Topological Conformal Field Theory

In this section we analyse the theory resulting from gauging the maximal diagonal group $G \subset G \times G$ [23]. Notice first of all, that the condition for the coset theory to exist, given by equation (5.2), is trivially satisfied in this case. From the explicit expression (5.6) for the coset energy-momentum tensor, it is clear that $T_{\mathfrak{g}/\mathfrak{g}}(z) = 0$, whence the total energy-momentum tensor $T(z)$ given by (5.18) agrees with $T'(z)$ given by (5.22). But, quite generally, we showed in the previous section that $T'(z)$ is BRST trivial, whence so is $T(z)$. Since $T(z)$ is the energy-momentum tensor of the theory—that is, it generates conformal transformations—the conformal Ward identities imply that correlation functions of BRST invariant operators will be insensitive to deformations in the metric; in particular, they will be (locally) constant. This means that the resulting $G/G$ coset theory is a topological conformal field theory (TCFT). As in every other known two-dimensional TCFT, we have in BRST cohomology the structure of a Batalin-Vilkowisky (BV) algebra. This is often induced from an $N=2$ superconformal algebra structure before descending to cohomology, but as shown by Getzler [26] in the case of $G/G$ ($G$ reductive), the BV structure is induced from a Kazama [27] algebra. For $G$ reductive, the existence of the Kazama algebra in the $G/G$ model was discovered independently by Isidro and Ramallo in [**28**]. In this section we will show that there is a Kazama algebra also in the self-dual case and that, provided that $L_0$ acts diagonally, the Kazama algebra induces a BV algebra structure on the TCFT.

The Kazama algebra

The above discussion and equation (5.30) show that the total energy momentum tensor of the $G/G$ gauged WZW model is BRST trivial. Let us introduce the following notation for some of the objects introduced in the previous section but in the case of $H = G$

$$\begin{aligned}\mathsf{G}^+(z) &= j_{BRST}(z) = (c^a(J_a + \tilde{J}_a + \tfrac{1}{2}J_a^{\text{gh}}))(z) \\ \mathsf{T}(z) &= T(z) = \Omega^{ab}(J_a J_b)(z) - \Omega^{ab}(\tilde{J}_a \tilde{J}_b)(z) - (b_a \partial c^a)(z) \\ \mathsf{G}^-(z) &= O(z) = \Omega^{ab} b_a \left(J_a - \tilde{J}_a\right)(z)\end{aligned} \qquad (6.1)$$

where $\Omega \equiv \Omega^{\mathfrak{g}}$, since there is no longer cause for confusion.

It is now a simple matter to compute the operator product algebra generated by the above fields. Let us proceed stepwise. First of all we notice that

$$\mathsf{G}^+(z)\mathsf{G}^+(w) = \text{reg}, \qquad (6.2)$$

a fact which is well-known for BRST differentials associated to affine Lie algebras (see for instance [**29**] which works out the semisimple case), but which also holds in the nonreductive case. Next we compute

$$\mathsf{G}^+(z)\mathsf{G}^-(w) = \frac{\dim \mathfrak{g}}{(z-w)^3} + \frac{\mathsf{J}(w)}{(z-w)^2} + \frac{\mathsf{T}(w)}{z-w} + \text{reg} , \qquad (6.3)$$

where $\mathsf{J}(z) = -(b_a c^a)(z)$ is the ghost current. It follows trivially that the following OPEs hold

$$\mathsf{J}(z)\mathsf{G}^\pm(w) = \frac{\pm \mathsf{G}^\pm(w)}{z-w} + \text{reg} . \qquad (6.4)$$

These OPEs are reminiscent of a twisted $N=2$ algebra, but that would require that $\mathsf{G}^-(z)\mathsf{G}^-(w)$ be regular. Instead one obtains

$$\mathsf{G}^-(z)\mathsf{G}^-(w) = \frac{-2\mathsf{F}(w)}{z-w} + \text{reg} , \qquad (6.5)$$

where we have introduced the field

$$\mathsf{F}(z) \equiv -\tfrac{1}{2}\kappa^{ab}(b_a \partial b_b)(z) - \tfrac{1}{2}f^{abc}(b_a b_b(J_c + \tilde{J}_c))(z) , \qquad (6.6)$$

where indices have been raised with $\Omega^{ab}$. The field $\mathsf{F}(z)$ is not just BRST

invariant but also BRST exact. Introducing the fermionic field

$$\Phi(z) = -\tfrac{1}{6} f^{abc} (b_a b_b b_c)(z) \ , \qquad (6.7)$$

we find that

$$\mathsf{G}^+(z)\Phi(w) = \frac{\mathsf{F}(w)}{z-w} + \mathrm{reg} \ . \qquad (6.8)$$

Noticing finally that

$$\mathsf{J}(z)\Phi(w) = \frac{-3\Phi(w)}{z-w} + \mathrm{reg} \ , \qquad (6.9)$$

puts us in a position to apply the theorem of Getzler [26], generalising a theorem of [29] and [30], which states that the fields $\mathsf{T}(z)$, $\mathsf{G}^\pm(z)$, $\mathsf{J}(z)$, $\Phi(z)$, and $\mathsf{F}(z)$ obey a Kazama algebra.

### The BV algebra structure of the $G/G$ TCFT

Throughout the remainder of this section we will assume that the zero mode $L_0$ of the topological energy-momentum tensor $\mathsf{T}(z)$ acts diagonally on the Hilbert space of the theory. That this assumption is not empty is evinced, for example, by theories with fields having logarithmic singularities in their OPEs (see for example, [31]). An equivalent assumption is that the Hilbert space of the theory decomposes into highest weight representations of the affine Lie algebra $\widehat{\mathfrak{g}}$ all of whose highest-weight vectors have a definite conformal weight. This assumption would not be necessary if we were considering a reductive Lie group $G$, but it is necessary in our more general case. We believe it to be a mild restriction.

It follows from associativity of the OPE that the OPE of BRST invariant fields is BRST invariant. This fact allows us to induce operations in BRST cohomology which, in the case of a TCFT, make it into a Batalin-Vilkowisky (BV) algebra.

By a BV algebra we mean a graded vector space $A = \bigoplus_p A^p$ with two operations:

$$\begin{aligned} \bullet &: A^p \otimes A^q \to A^{p+q} \qquad (a,b) \mapsto ab \\ \Delta &: A^p \to A^{p-1} \end{aligned} \qquad (6.10)$$

enjoying the following two properties: $\Delta^2 = 0$, and $\bullet$ is an associative, supercommutative multiplication:

$$a(bc) = (ab)c \qquad \mathrm{and} \qquad ab = (-)^{pq} ba \quad \mathrm{for} \quad a \in A^p, \ b \in A^q \ . \qquad (6.11)$$

These two operations allow us to define a bracket as follows

$$[a,b] = (-)^p \left( \Delta(ab) - (\Delta a)b - (-)^p a \Delta b \right) \ , \qquad (6.12)$$

for all $a \in A^p$; in other words, the bracket measures the failure of $\Delta$ of being

– 26 –

a derivation over •. It is not difficult to show that the bracket (6.12) is super-antisymmetric, but where the degree is shifted by one from the natural one; that is,

$$[a, b] = -(-)^{(p+1)(q+1)}[b, a] \quad \text{for} \quad a \in A^p, \text{ and } b \in A^q . \tag{6.13}$$

Similarly one can show that for any $a \in A^p$, $[a, -]$ is a derivation of degree $p + 1$ over •:

$$[a, bc] = [a, b]c + (-)^{(p+1)q}b[a, c] \quad \text{for } b \in A^q, \text{ and } c \in A . \tag{6.14}$$

Finally one proves that $[-, -]$ obeys the Jacobi identity in such a way that it makes $\widehat{A} = \bigoplus_p \widehat{A}^p$ where $\widehat{A}^p = A^{p-1}$ into a Lie superalgebra.

Let us remark that a graded Lie algebra $A$ with operations • and $[-, -]$ satisfying the above properties is called a Gerstenhaber algebra. If an operator $\Delta : A^p \to A^{p-1}$ as above exists in such a way that equation (6.12) is satisfied, then one says that $\Delta$ generates the bracket $[-, -]$. The properties of the bracket do not force $\Delta^2$ to vanish, but simply to be of sufficiently low "order" (see [**32**] for an insightful discussion). We are not aware of any example of a Gerstenhaber algebra whose bracket is generated by a $\Delta$ which cannot be chosen to square to zero. Notice that $\Delta$ is defined up to the addition of a (anti)derivation.

BV algebras are common objects: the BRST cohomology of *any* string theory is a BV algebra [**33**], as is the (anti)chiral ring of any $N=2$ superconformal field theory. In fact, as we now show, TCFTs obtained from Kazama algebras are also BV algebras. To do this we will need to define the operations of • and $\Delta$ on the cohomology of the BRST operator, which we take to be the zero mode of $\mathsf{G}^+(z)$ as in the $G/G$ example above.

We find it useful to employ the following notation: if $A$ and $B$ are fields in any CFT, we write their OPE as

$$A(z)B(w) = \sum_{n \ll \infty} \frac{[A, B]_n(w)}{(z - w)^n} , \tag{6.15}$$

which define the brackets $[-, -]_n$. The commutativity and associativity properties of the OPE imply certain identities obeyed by these brackets, and are in fact equivalent to them. The first identity we will need is the commutativity

of the OPE:

$$[A, B]_n - (-)^{|A||B|+n}[B, A]_n = \sum_{\ell \geq 1} \frac{(-)^{1+\ell}}{\ell!} \partial^\ell [A, B]_{n+\ell}$$
$$= (-)^{|A||B|} \sum_{\ell \geq 1} \frac{(-)^{n+\ell}}{\ell!} \partial^\ell [B, A]_{n+\ell} .$$
(6.16)

The second identity is a Jacobi-like identity which follows from the associativity of the OPE:

$$[A, [B, C]_n]_{m>0} = (-)^{|A||B|} [B, [A, C]_m]_n + \sum_{\ell=0}^{m-1} \binom{m-1}{\ell} [[A, B]_{m-\ell}, C]_{n+\ell} .$$
(6.17)

Notice that, in particular, for any $A(z)$, the operation $[A, -]_1$ is a derivation over all the $[-, -]_n$. Finally we notice that the bracket $[-, -]_0$ coincides with the normal-ordered product, so that we will abbreviate $[A, B]_0$ simply by $(AB)$. The normal-ordered product is neither associative nor commutative; but rather obeys the following "rearrangement lemma"

$$(A(BC)) - (-)^{|A||B|}(B(AC)) = ((AB)C) - (-)^{|A||B|}((BA)C) . \quad (6.18)$$

Finally notice the following properties of the derivative, which in particular allow us to compute the brackets $[-, -]_{n<0}$ in terms of the others

$$[\partial A, B]_n = (1 - n)[A, B]_{n-1} \quad \text{and} \quad [A, \partial B]_n = (n - 1)[A, B]_{n-1} + \partial [A, B]_n .$$
(6.19)

Notice that the BRST operator acts on fields as $[\mathsf{G}^+, -]_1$, which by (6.17) is a derivation. This means that if $A$ and $B$ are BRST invariant fields, so is $[A, B]_n$ for any $n$. Moreover, if either $A$ or $B$ is BRST trivial, then so is $[A, B]_n$ for any $n$. Therefore the BRST cohomology inherits the brackets $[-, -]_n$, which still obey the above properties. We define $H = \bigoplus_q H^q$ to be the BRST cohomology (on fields), where $q$ refers to the $U(1)$-charge defined as the eigenvalue of the derivation $[\mathsf{J}, -]_1$; in the $G/G$ model it agrees with the ghost number. We will show that $H$ is a BV algebra. In order to do this we need to define the two operations: $\bullet$ and $\Delta$. We define $\bullet$ as the operation induced from the normal-ordered product. Because $\mathsf{T}(z)$ is BRST trivial, all the operations $[\mathsf{T}, -]_n$ are zero in cohomology. In particular, $\partial = [\mathsf{T}, -]_1$ is zero. (This explains why the correlation functions in a TCFT are constant.) From (6.16) it follows that the normal-ordered product is supercommutative in cohomology, since the RHS is a total derivative of BRST invariant fields, and hence is BRST trivial. Similarly, it is not hard to show from the rearrangement lemma (6.18), that if the normal-ordered product is supercommutative it is also associative. This shows that $\bullet : H^p \otimes H^q \to H^{p+q}$ obeys the right axioms.

The definition of $\Delta$ is trickier. Because $L_0 = [\mathsf{T}, -]_2$ is BRST trivial and (by hypothesis) acts diagonally, a standard argument shows that all the cohomology resides at the $L_0 = 0$ eigenspace. That is, any BRST invariant field of nonzero conformal weight is BRST trivial. This means that we do not lose any cohomology by restricting ourselves from the start to fields of zero conformal weight. Equivalently, if we are not prepared to restrict ourselves to this subspace, what this tells us is that every BRST cohomology class always has one representative of zero conformal weight. From now on we will always assume that every BRST invariant field that we write down has zero conformal weight. Now we can define $\Delta$ as $[\mathsf{G}^-, -]_2$. It is easy to verify that it sends BRST invariant fields of zero conformal weight to themselves. Moreover, although $\Delta^2 \neq 0$, it is zero in cohomology. In fact, if we let $d = [\mathsf{G}^+, -]_1$ denote the BRST differential and $\delta = [\Phi, -]_3$, then one finds (from (6.17)) that

$$\Delta^2 = -[d, \delta] \ , \tag{6.20}$$

which says that $\Delta^2$ is chain homotopic to zero, and hence zero in cohomology. In other words, $\Delta : H^q \to H^{q-1}$ obeys the right axioms for $(H, \bullet, \Delta)$ to be a BV algebra. Notice that using (6.12) and (6.17) that the bracket can be rewritten as

$$[A, B] = (-)^{|A|}[[\mathsf{G}^-, A]_1, B]_1 \ . \tag{6.21}$$

We would like to emphasise that provided that $L_0 = [\mathsf{T}, -]_2$ acts diagonally, any Kazama algebra gives rise to a BV algebra in cohomology; so that the above results are more general than the $G/G$ model (see [26] for Kazama algebras obtained from reductive Manin pairs).

§7  CONCLUSIONS

Let us summarise our main results. We have studied the general two-dimensional WZW model with target space a nonreductive Lie group $G$ and have proven that the resulting quantum field theory is nonperturbatively conformally invariant by constructing a Sugawara tensor. The existence of this more general Sugawara construction was already known [11] [13], as well as the existence of the WZW model [11]; both the existence of both construction at the same time require that the Lie group possess two bi-invariant metrics related by a shift. What we have shown in this paper is that these two nondegeneracy conditions on the metrics can be simultaneously satisfied. This result was obtained after a careful analysis at the structure of self-dual Lie algebras.

We then turned our attention to the gauging of the WZW model. After recalling the obstructions which prevent us from gauging a subgroup $H \subset G \times G$ of the isometry group, we specialised to the case of a diagonal subgroup, for which the obstruction vanishes. We analysed the quantum theory resulting from a diagonal gauging and we proved that it yields a conformal field theory whose energy-momentum tensor agrees with that of a coset construction. Strictly speaking, the gauged WZW model is conformal precisely when a condition is met which agrees with the condition guaranteeing the existence of the coset construction.

Finally we studied the topological conformal field theory resulting from gauging the maximal diagonal subgroup $G \subset G \times G$. The resulting $G/G$ theory is a TCFT with a BV algebra structure in cohomology which comes induced from a Kazama algebra present in the BRST complex.

There are a few immediate extensions of the results in this paper. One obvious extension is to gauge other anomaly-free subgroups and see what other conformal field theories such gaugings might give rise to. In particular it would be interesting to generalise the Drinfel'd–Sokolov construction to these more general WZW models. At the moment we have little to say on these matters except for the fact that taking into account the condition (4.2) to be able to gauge a subgroup, we notice that a nonreductive self-dual Lie algebra has generically more subalgebras satisfying this condition than their reductive counterparts.

Work is in progress on the supersymmetrisation of this construction to $N{=}1$ supersymmetric WZW models with the ultimate aim to generalise the Kazama-Suzuki construction. As discussed in [**11**], the $N{=}1$ Sugawara construction exists if and only if the $N{=}0$ Sugawara construction exists. Similarly, our results of Section 3 imply that the correspondence between the $N{=}1$ WZW model and the $N{=}1$ Sugawara construction survives the generalisation.

Another possible supersymmetric extension would be to consider WZW models based on Lie supergroups; or at least current algebras defined from affine Lie superalgebras. Many of the results in this paper generalise to the case of Lie superalgebra, for example the Sugawara construction, as we now briefly recall. Similar results were obtained independently in [**34**].

Let $\mathfrak{g} = \mathfrak{g}_0 \oplus \mathfrak{g}_1$ be a Lie superalgebra with homogenoeus basis $\{J_A\}$. Let $\langle - , - \rangle$ denote a symmetric bilinear form on $\mathfrak{g}$ and let $g_{AB} = \langle J_A , J_B \rangle$. In a $\mathbb{Z}_2$-graded category, "symmetry" means $g_{AB} = (-)^{AB} g_{BA}$. Finally let us introduce the structure constants $f_{AB}{}^C$ by $[J_A, J_B] = f_{AB}{}^C J_C$. They are antisymmetric; that is, $f_{AB}{}^C = -(-)^{AB} f_{BA}{}^C$. Moreover they obey the (super)Jacobi identity:

$$f_{BC}{}^D f_{AD}{}^E = f_{AB}{}^D f_{DC}{}^E + (-)^{AB} f_{AC}{}^D f_{BD}{}^E \ . \tag{7.1}$$

The affine Lie superalgebra $\widehat{\mathfrak{g}}$ is defined by the following OPE:

$$J_A(z)J_B(w) = \frac{g_{AB}}{(z-w)^2} + \frac{f_{AB}{}^C J_C(w)}{z-w} + \text{reg} . \qquad (7.2)$$

Given (7.1), the mode algebra defined by this OPE obeys the Jacobi identity provided that $g_{AB}$ is ad-invariant:

$$f_{AB}{}^D g_{DC} = f_{BC}{}^D g_{AD} . \qquad (7.3)$$

Notice that the bilinear form $g_{AB}$ is forced to be even; that is, $\langle \mathfrak{g}_0 , \mathfrak{g}_1 \rangle = 0$.

Now let $\Omega^{AB} = (-)^{AB}\Omega^{BA}$ be a symmetric even bivector and define $T = \Omega^{AB}(J_A J_B)$. We demand that $T$ obey the Virasoro algebra and that every $J_A(z)$ be a primary field of weight one. As in the nonsupersymmetric case the latter condition implies the former. The condition for $J_A(z)$ to be a weight 1 primary is given by

$$J_A(z)T(w) = \frac{J_A(w)}{(z-w)^2} + \text{reg} . \qquad (7.4)$$

Computing the LHS we have three poles. The third order pole vanishes by (7.3) and the symmetry of $\Omega^{AB}$. The second order pole gives the following equation

$$(-)^{AE}\Omega^{EC}g_{AC} + \Omega^{BC}f_{AB}{}^D f_{DC}{}^E + g_{AB}\Omega^{BE} = \delta_A{}^E ; \qquad (7.5)$$

whereas the first order pole simply states the ad-invariance of the bivector $\Omega^{AB} J_A \otimes J_B$; that is,

$$\Omega^{AE} f_{CA}{}^D + (-)^{CD}\Omega^{DB} f_{CB}{}^E = 0 . \qquad (7.6)$$

Using (7.6), we can rewrite (7.5) as follows:

$$(2g_{AC} + \kappa_{AC})\Omega^{CE} = \delta_A{}^E , \qquad (7.7)$$

where we have introduced the Killing form

$$\kappa_{AC} \equiv \text{Str ad } J_A \text{ ad } J_C = (-)^D f_{AB}{}^D f_{CD}{}^B . \qquad (7.8)$$

Equation (7.7) clearly says that $\Omega^{AB}$ is invertible with inverse

$$\Omega_{AB} = 2g_{AB} + \kappa_{AB} . \qquad (7.9)$$

In summary, provided that the currents $\{J_A(z)\}$ obey (7.2) with $g_{AB} = \frac{1}{2}(\Omega_{AB} - \kappa_{AB})$, and $\kappa_{AB}$ given by (7.8), then $T = \Omega^{AB}(J_A J_B)$ will obey the Virasoro algebra with central charge $c$ given by

$$c = \Omega^{AB}\Omega_{AB} - \Omega^{AB}\kappa_{AB} = \dim \mathfrak{g}_0 - \dim \mathfrak{g}_1 - \Omega^{AB}\kappa_{AB} . \qquad (7.10)$$

In addition, the currents $\{J_A(z)\}$ are conformal primaries of weight 1.

Thus as in the nonsupersymmetric case, a Sugawara construction exists for any self-dual Lie superalgebra. The problem remains to characterise the self-dual Lie superalgebras. Unlike self-dual Lie algebras, there does not exist—to our knowledge—a structure theorem à la Medina–Revoy. In fact, a careful look at the theorem shows that almost everything works, except for one technical result on the splitting of exact sequences involving simple Lie algebras (see the appendix of [**19**]). In particular the notion of a double extension still works in this case and one can construct self-dual Lie superalgebras in this fashion, but there is no guarantee that all self-dual Lie superalgebras are obtained this way. The generalisation of the theorem of Medina and Revoy to category of Lie superalgebras is, from our point of view, a very interesting open problem.

## ACKNOWLEDGEMENTS

We would like to acknowledge useful conversations with Takashi Kimura and George Thompson. We would also like to thank Francisco Figueirido for his beautiful (unpublished) notes on his lectures on anomaly computations and the WZW model given at Stony Brook in the late 80's.